\documentclass[aps,preprint,amssymb,12pt,floatfix]{revtex4}
\usepackage{subfigure}
\usepackage{graphicx}
\usepackage{rotating}

\begin{document}
\title{Mechanical unfolding of RNA hairpins}
\author{Changbong Hyeon$^1$ and D. Thirumalai$^{1,2}$}
\thanks{Corresponding author phone: 301-405-4803; fax: 301-314-9404; thirum@glue.umd.edu}
\affiliation{$^1$Biophysics Program,
Institute for Physical Science and Technology\\
$^2$Department of Chemistry and Biochemistry\\
University of Maryland, College Park, MD 20742\\
\[
\]
\noindent 23 pages, 6 figures, 245 words in abstract, 46757 characters
}

\date{\small \today}

\baselineskip = 22pt
\begin{abstract}
Mechanical unfolding trajectories, generated by applying constant force in
optical tweezer experiments, show that RNA hairpins and the P5abc
subdomain of the group I intron unfold reversibly. We use coarse-grained Go-like
models for RNA hairpins to explore forced-unfolding over a broad range of
temperatures. A number of predictions that are amenable to experimental
tests are made. 
At the critical force
the hairpin jumps between folded and unfolded conformations without
populating any discernible intermediates.  
The phase diagram in the force-temperature ($f,T$) plane
shows that the hairpin unfolds by an all-or-none process. The cooperativity
of the unfolding transition increases dramatically at low
temperatures. Free energy of stability, obtained from time averages of
mechanical unfolding trajectories, coincide with ensemble averages which
establishes  ergodicity. The hopping time between the the native basin of
attraction (NBA) and the unfolded basin increases dramatically along
the phase boundary. Thermal unfolding is stochastic whereas mechanical
unfolding occurs in \emph{quantized steps} with great variations in the
step lengths. Refolding times, upon force quench, from stretched states to
the NBA is \emph{at least an order of magnitude} greater than folding times by
temperature quench. Upon force quench from stretched states the NBA is
reached in at least three stages. In the initial stages the mean
end-to-end distance decreases nearly continuously and only in the last stage
there is a sudden transition to the NBA.
Because of the generality of the results we propose that similar behavior should be 
observed in force quench refolding of proteins.
\end{abstract}
\maketitle

\newpage
\section{Introduction}
Unraveling the complexity of the energy landscape of RNA molecules requires exploration of their assembly and 
unfolding over a wide range of external conditions. 
In the last decade a combination of experiments, theoretical arguments, and simulations have been used 
to decipher the folding mechanisms of RNA molecules \cite{OnoaCOSB04,TreiberCOSB01,ThirumARPC01}. 
These studies have shown that RNA folding depends critically on a number of factors
including valence and shape of counterions \cite{KoculiJMB04}, and temperature.
Somewhat more surprisingly recent experiments have shown that the folding mechanisms depend sensitively 
on the initial folding conditions \cite{RussellPNAS02}. 
In conventional experiments the difficult-to-characterize unfolded conformations are typically generated 
by altering temperature or by lowering the counterion concentration. 
In contrast, well-defined and vastly different initial conditions can be realized by applying force. 
Indeed, in remarkable experiments Bustamante and coworkers \cite{Bustamante2,Bustamante3} have generated mechanical unfolding 
trajectories for RNA hairpins and \emph{T. thermophila} ribozyme.
These experiments, which use constant external force to denature folded RNA, show that unfolding involves multiple routes 
in which a number of kinetic intermediates are sampled in the transition from the folded state to 
a stretched conformation \cite{Bustamante2, Bustamante3}. 
The lifetimes of the intermediates vary considerably, which is indicative of the large dispersion in the unfolding pathways. 
Thus, force unfolding is a powerful method to probe, at the single molecule level, regions of the energy landscape that are \emph{inaccessible in conventional folding experiments}. 
In addition, to the importance of these experiments to map the RNA folding landscape 
response of RNA to locally applied force may also be relevant in understanding cellular processes such as 
mRNA translocation through ribosomes, viral replication, and enzymatic activity of RNA dependent RNA polymerases. 

In the force-induced unfolding experiments 
mechanical force, $f$, was applied using optical tweezers either to a part or to 
the whole \emph{Tetrahymena} ribozyme assembly in differing ionic conditions. 
In their first report Liphardt \emph{et al.} \cite{Bustamante2} showed that a simple hairpin, three helix junction, and the 
P5abc subdomain of the \emph{Tetrahymena} ribozyme can fold reversibly when subject to a constant force. 
At the transition force the systems hop between folded and unfolded states. 
Assuming that the system is ergodic the dynamics of the reversible folding 
was used to calculate force-dependent equilibrium properties of the RNA constructs. 
These experiments established that $f$ $-$a new variable to initiate unfolding$-$
is a viable way to measure free energy difference between 
folded and unfolded states and to locate transition states with the mean extension of the molecule as a reaction coordinate.

Mechanical unfolding experiments on RNA have already lead to a number of theoretical studies \cite{MezardEPJE02,MarkoEPJE03,HwaBP01,HwaBP03}
that have addressed different 
aspects of forced-unfolding. 
Inspired by these experiments and building on previous theoretical works 
we report here the results for forced-unfolding of a small RNA hairpin 
using coarse-grained off-lattice simulations under varying forces and temperatures.
We choose small hairpins for the preliminary study 
because they undergo reversible folding under force and they represent a basic subunit of large RNA assemblies. 

We address the following questions:
(1) What are the forced-unfolding pathways and how they differ from thermal denaturation? 
(2) How do the diagram of states change as $T$ and $f$ are varied?
(3) What are the differences in the time scales and pathways in force quench refolding and thermal refolding?
We find that, just as in proteins \cite{KlimovPNAS99}, forced-unfolding occurs in quantized steps whereas the thermal unfolding is 
stochastic. 
Even for the simple hairpin we find a well-defined equilibrium phase diagram in the ($f,T$) plane in which 
hairpin states are separated by a phase boundary from the unfolded states. 
Surprisingly, when refolding is initiated by quenching to zero force from high forces, the folding 
occurs in \emph{multiple stages with the initial compaction being nearly continuous}. 
Remarkably, the refolding under force quench is nearly an order of magnitude greater than thermal refolding time.\\

\section{Methods}

{\bf Hairpin sequence:} We have studied the thermal and forced-unfolding of a 22-nucleotide hairpin, P5GA, 
that is similar to P5ab in the P5abc domain of group I intron.
Both these structures have GA mismatches and are characterized by the presence of GAAA tetraloop. 
The sequence of P5GA is GGCGAAGUCGAAAGAUGGCGCC
and its NMR structure has been determined \cite{TinocoJMB2000} (PDB id:1eor). \\

{\bf Model:}
Building on our previous studies on proteins \cite{Klimov2}
we introduce a coarse-grained off-lattice model of RNA by representing each nucleotide
by three beads with interaction sites corresponding to phosphate group (P),
ribose group (S), and the base (B) (Fig.\ref{coarse_nucl}-A).
In this model the RNA backbone is reduced to the polymeric structure $-(P-S)_n-$ and 
the base is covalently linked to the ribose center. 
Thus, a RNA molecule with N nucleotides corresponds to 3N interaction centers.
The potential energy of a conformation is written as 
$V_{TOT}=V_{BL}+V_{BA}+V_{DIH}+V_{STACK}+V_{NON}+V_{ELEC}$ where $V_{BL}$ the stretching potential between covalently connected moieties accounts for chain connectivity. 
The angular degrees of freedom are described by the bond angle potential, $V_{BA}$, and the dihedral angle
term $V_{DIH}$ \cite{KlimovFoldDes98}. In this paper we use a Go model in which interactions in the native structure are attractive 
while all other interactions are repulsive. 

Simple RNA secondary structures are stabilized largely by stacking interactions
whose context dependent values are known \cite{WalterPNAS94,MathewsJMB99}.
In the native state the P5GA hairpin has nine hydrogen bonds between 
the base pairs including two GA mismatch pairs \cite{TinocoJMB2000}. 
The stacking interactions that stabilize a hairpin is $V_{STACK}=\sum_{i=1}^{n_{max}}V_i$ where $n_{max}=8$ in P5GA. 
The orientational dependent terms $V_i$ is taken to be 
\begin{eqnarray}
V_i(\{\phi\},\{\psi\},\{r\};T)=\Delta G_i(T)&\times&e^{-\alpha_{st}\{sin^2(\phi_{1i}-\phi_{1i}^{o})+sin^2(\phi_{2i}-\phi_{2i}^{o})+sin^2(\phi_{3i}-\phi_{3i}^{o})+sin^2(\phi_{4i}-\phi_{4i}^{o})\}}\nonumber\\
&\times&e^{-\beta_{st}\{(r_{ij}-r_{1i}^{o})^2+(r_{i+1j-1}-r_{2i}^{o})^2\}}\nonumber\\
&\times&e^{-\gamma_{st}\{sin^2(\psi_{1i}-\psi_{1i}^{o})+sin^2(\psi_{2i}-\psi_{2i}^{o})\}}
\end{eqnarray}
where $\Delta G(T)=\Delta H-T\Delta S$, the bond angles $\{\phi\}$ are
$\phi_{1i}\equiv\angle S_iB_iB_j$, $\phi_{2i}\equiv\angle B_iB_jS_j$, 
$\phi_{3i}\equiv\angle S_{i+1}B_{i+1}B_{j-1}$, $\phi_{4i}\equiv\angle B_{i+1}B_{j-1}S_{j-1}$, 
the distance between two paired bases $r_{ij}=|B_i-B_j|$, $r_{i+1j-1}=|B_{i+1}-B_{j-1}|$, and
$\psi_{1i}$ and $\psi_{2i}$ are the dihedral angles formed by the four beads 
$B_iS_iS_{i+1}B_{i+1}$ and $B_{j-1}S_{j-1}S_jB_j$, respectively. 
The superscript $o$ refers to angles and distances in the PDB structure. 
The values of $\alpha_{st}$, $\beta_{st}$ and $\gamma_{st}$ are 
1.0, 0.3\AA$^{-2}$ and 1.0 respectively. 
We take $\Delta H$ and $\Delta S$ from Turner's thermodynamic 
data set \cite{MathewsJMB99,WalterPNAS94}.
There are no estimates for GA related stacking interactions, 
which typically do not form a stable bond and hence is considered a mismatch.
Because of the absence of stacking parameters for the GA pair, 
we use the energy associated with GU in place of GA. 

To mimic the hydrophobicity of purine/pyrimidine group, we use the Lennard-Jones (LJ)
interactions between non-bonded interaction centers.
The total nonbonded potential is 
\begin{equation}
V_{NON}=\sum_{i=1}^{N-1}\sum_{j=i+1}^NV_{B_iB_j}(r)+\sum_{i=1}^N\sum_{m=1}^{2N-1}{'}V_{B_i(SP)_m}(r)+\sum_{m=1}^{2N-4}\sum_{n=m+3}^{2N-1}V_{(SP)_m(SP)_n}(r)
\label{eqn:V_non}
\end{equation} 
where $r=|\vec{r}_i-\vec{r}_j|$,
the prime in the second term on the Eq.(\ref{eqn:V_non}) denotes the condition $m\neq2i-1$, and 
$(SP)_m=S_m$ or $P_m$ depending on index $m$. 
A native contact exists between two non-covalently bound beads provided they are within 
a cut-off distance $r_c$ (=7.0\AA).
Two beads beyond $r_c$ are considered to be non-native.
For a native contact, 
\begin{equation}
V_{\xi_i\eta_j}(r)=C_h^{\xi_i\eta_j}[(\frac{r^o_{ij}}{r})^{12}-2(\frac{r^o_{ij}}{r})^6]
\end{equation}
where $r^o_{ij}$ is the distance between beads in PDB structure and $C_h^{\xi_i\eta_j}=1.8kcal/mol$ for all native contact pairs except 
for $B_{10}B_{13}$ base pair associated with the formation of the hairpin loop, 
for which $C_h^{B_{10}B_{13}}=3.0kcal/mol$. 
The additional stability for the base pair associated with loop formation is similar to the 
Turner's thermodynamic rule for the free energy gain in the tetraloop region.
For beads beyond $r_c$ the interaction is
\begin{equation}
V_{\xi_i\eta_j}(r)=C_R [(\frac{a}{r})^{12}+(\frac{a}{r})^6]
\end{equation}
with $a=3.4$\AA\ and $C_R=1kcal/mol$. 
The value of $C_h$ has been chosen so that the hairpin undergoes a first order transition from unfolded states. Our results are not sensitive to minor variations in $C_h$. 

The electrostatic potential between the phosphate groups is assumed to be pairwise additive  
$V_{ELEC}=\sum_{i=1}^{N-1}\sum_{j=i+1}^NV_{P_iP_j}(r)$.
For $V_{P_iP_j}(r)$ we assume Debye-H\"{u}ckel interaction, which accounts for screening by condensed 
counterions and hydration effects, and is given by 
\begin{equation}
V_{P_iP_j}=\frac{z_{P_i}z_{P_j}e^2}{4\pi\epsilon_0\epsilon_r r}e^{-r/\l_D}
\label{eqn:electrostatic}
\end{equation}
where $z_{P_i}=-1$ is the charge on the phosphate ion, $\epsilon_r=\epsilon/\epsilon_0$ 
and the Debye length $l_D=\sqrt{\frac{\epsilon_rk_BT}{8\pi k_{elec}e^2I}}$ with $k_{elec}=\frac{1}{4\pi\epsilon_0}=8.99\times 10^{9}JmC^{-2}$.
To calculate the ionic strength $I=1/2\sum_iz_i^2c_i$, 
we use the value $c_i=200mM$-$NaCl$ from the header of PDB file \cite{TinocoJMB2000}. 
We use $\epsilon_r=10$ in the simulation \cite{MisraPNAS01}.
Because the Debye screening length $\sim\sqrt{T}$ the strength of electrostatic interaction between the phosphate group is 
temperature dependent even when we ignore the variations of $\epsilon$ with $T$.
At room temperature ($T\sim 300K$) 
the electrostatic repulsion between the phosphate groups at $r\sim$5.8\AA, 
which is the closest distance between phosphate groups, is $V_{P_iP_{i+1}}\sim 0.5kcal/mol$.
It follows that $V_{ELEC}$ between phosphate groups across the base pairing ($r=16\sim18\AA$) 
is almost negligible.\\

{\bf Simulations:}
The dynamics of stretching is obtained by integrating the Langevin equation.
Forced-unfolding simulations are performed by applying a constant force to the $S$ bead at one end of the molecule.
Using a typical value for the mass of a bead in a nucleotide ($B_i$, $S_i$ or $P_i$), 
$m$, $100g/mol\sim160g/mol$,
the average distance between the adjacent beads $a=4.6$\AA, 
the energy scale $\epsilon_h=1\sim2kcal/mol$, the natural time is $\tau_L=(\frac{ma^2}{\epsilon_h})^{1/2}=1.6\sim2.8ps$.
We use $\tau_L=2.0ps$ to convert simulation times into real times.
To estimate the time scale for thermal and mechanical unfolding dynamics we use a Brownian dynamics algorithm \cite{McCammonJCP78,KlimovFoldDes98} 
for which the natural time for the overdamped motion is $\tau_H=\frac{\zeta\epsilon_h}{T}\tau_L$. 
We used $\zeta=50\tau_L^{-1}$ in the overdamped limit, that approximately corresponds to friction constant in water.
At $T=290K$, $10^6$ time steps correspond to $3.5\mu s$.
To probe the thermodynamics and kinetics of folding we used a number of physical quantities (end-to-end distance ($R$),
fraction of native contacts ($Q$), structural overlap function ($\chi$), number of hydrogen bonds $n_{bond}$, etc) 
to monitor the structural change in the hairpin. 
The free energy profiles and the phase diagram were obtained using an adaptation of 
multiple histogram method \cite{KumarJCC1992} for force unfolding of biomolecules (CH and DT, unpublished). \\ 
 
\section{Results and Discussion}

{\bf Determination of the Native state:}
Using a combination of multiple slow cooling, simulated annealing, and steepest descent quenches we determined the native 
structure of the hairpin.
To ensure that there is no other structure with lower energy, the structure obtained 
from steepest descent method is reheated to $T=100K$ and cooled down again. 
By repeating this process we obtained the computed native conformation which has a 
RMSD of 0.1\AA\ with respect to the PDB structure..
The bulk of the contribution to the total energy, 
$V_{TOT}=-154kcal/mol$, of the native conformation arises from $V_{STACK}=-95.5kcal/mol$, $V_{NON}=-59.2kcal/mol$.\\

{\bf Force-temperature ($f$,$T$) phase diagram:}
The diagram of states in the ($f,T$) plane shows that P5GA hairpin behaves 
as a ``two-state'' folder (Fig.\ref{phase}).
In the absence of force ($f=0$ $pN$) the folding/unfolding
transition midpoint is at $T_m=341K$ using 
$\langle Q\rangle$ as an order parameter.
At $T=290K$ the equilibrium force, required to unfold the P5GA is about $7pN$ (Fig.\ref{phase}), 
which is half the value for unfolding P5ab. The difference is, in all likelihood, 
due to the smaller length of P5GA. 
As force increases, $T_F$ decreases monotonically, so that the transition midpoints 
($T_m,f_m$) form a phase boundary separating the folded ($\langle Q\rangle>0.5$ 
and $\langle R\rangle<3 nm$) and the unfolded states.
The phase boundary is sharp at low $T_m$ and large $f_m$, 
but is fuzzy when the force is weak. 
The locus of points separating the unfolded and folded states can be fit using
\begin{equation}
f_c\sim f_o\left(1-\left(\frac{T}{T_m}\right)^{\alpha}\right)
\label{eqn:boundaryfit}
\end{equation}
where $f_o$ is the critical force at the low temperature and $\alpha$(=6.4) is a sequence dependent exponent.
The large value of $\alpha$ is indicative of a weak first order transition separating the hairpin and unfolded states \cite{KlimovPNAS99}.
\\

{\bf Two state dynamics and equilibrium:}
We used the thermodynamic relation $\log{K_{eq}(f)}=-\Delta F_{UF}/k_BT+f\cdot\Delta x_{UF}/k_BT$ and
the dependence of $\log{K_{eq}}$ ($K_{eq}$ is computed as time averages of the traces in Fig.\ref{compKeq}-A) on $f$ 
to estimate $\Delta F_{UF}$ and $\Delta x_{UF}$ which is 
the equilibrium distance separating the native basin of attraction (NBA) 
and the basin corresponding to the ensemble of unfolded states (UBA). 
The transition midpoint ($K(f_m)=1$) 
gives $f_m\approx 6pN$ in excellent agreement with the value obtained from the equilibrium phase diagram (Fig.\ref{phase}-A) which establishes ergodicity.
From the slope, $\frac{\partial\log{K_{eq}(f)}}{\partial f}=1.79pN^{-1}$, 
$\Delta x_{UF}\approx7.5nm$ we found, by extrapolation to $f=0$, that $\Delta F_{UF}\approx6.2kcal/mol$ 
under the assumption that $\Delta x_{UF}$ is \emph{constant and independent of f}.

The independence of $\Delta x_{UF}$ on $f$ was also used by Liphardt \emph{et al.} \cite{Bustamante2} to estimate $\Delta F_{UF}$. 
To check the validity of this assumption we computed free energy profiles using the multiple
histogram method with $R$ as the progress variable.
At $T=305K$, we find, from the free energy profile $F(R)$, 
that $\Delta F_{UF}\approx 5.8kcal/mol$ and $\Delta x_{UF}\approx 5.2nm$. 
Although the change in $\Delta F_{UF}$ computed from estimate of $K_{eq}(f)$ based on hopping dynamics 
and the ``exact'' result is small ($\approx$ 7\%) there is substantital difference in $\Delta x_{UF}$.
The exact free energy profile (Fig. \ref{compKeq}-C) clearly shows that 
\emph{$\Delta x_{UF}$ varies with $f$} because of large variations in the unfolded states. 
In general the assumption that $\Delta x_{UF}$ is a constant leads to an overestimate of both $\Delta F_{UF}$
and $\Delta x_{UF}$.\\

{\bf Cooperativity of unfolding depends on force:}
Slice of the phase diagram at either constant $f$ or constant $T$
shows the typical sigmoidal curves for $\langle Q\rangle$ as a function of  
either $f$ or $T$ (Fig.\ref{coop}).
The cooperativity of the transition depends on whether $T$ or $f$ is held constant.
We use the dimensionless cooperativity index $(\Omega_c^T)_f$ with respect to $T$ \cite{ThirumFoldDes98}. 
\begin{equation}
(\Omega_c^T)_f=\frac{T_{max}^2}{\Delta T}\vert\frac{d\langle Q\rangle}{dT}\vert_{max}
\end{equation}
where $\Delta T$ is the full width at the half maximum of $|(\frac{d\langle Q\rangle}{dT})_{max}|$
and $T_{max}$ is the temperature at which $\frac{d\langle Q\rangle}{dT}$ has a maximum.
Similarly, the dimensionless cooperativity index with respect to $f$ can be defined.
The force dependent cooperativity index $(\Omega_c^T)_f$ 
has a maximum around $f\sim 10pN$, whereas
$(\Omega_c^f)_T$ decreases monotonically to zero as $T$ increases (Fig.\ref{coop}-B and \ref{coop}-D). 
The difference between $(\Omega^T_c)_f$ and $(\Omega^f_c)_T$ arises because thermal denaturation at all forces is 
more stochastic while forced-unfolding disrupts RNA structures in steps.\\

{\bf Time scales of hopping transition:}
In the RNA pulling experiments \cite{Bustamante2} the time interval between the 
hopping transitions between folded and unfolded states at midpoint of force was measured at a single temperature. 
We have evaluated the dynamics along the phase boundary ($T_m$,$f_m$) (Fig.\ref{barrier_and_time})
to evaluate the variations in the free energy profiles and the dynamics of transition from the 
NBA to UBA. 
Along the boundary ($T_m$,$f_m$) there are substantial changes in the free energy landscape (Fig.\ref{barrier_and_time}-A).
The free energy barrier $\Delta F^{\ddagger}$ increases dramatically at low $T$ and high $f$.
We predict that the weakly first order phase transition at $T\approx T_m$ and low $f$ becomes increasingly stronger 
as we move along the ($T_m$,$f_m$) boundary to low $T$ and high $f$.

The two basins of attraction (NBA and UBA) are separated by a free energy barrier whose height increases 
as force increases (or temperature decreases) along ($T_m$,$f_m$) (Fig.\ref{barrier_and_time}-A). 
The hopping time $\tau_h$ along ($T_m$,$f_m$) is 
\begin{equation}
\tau_h=\tau_0\exp{(\Delta F^{\ddagger}/k_BT)}. 
\end{equation}
To estimate the variations in $\tau_h$ along the ($T_m$,$f_m$) boundary, 
we performed three very long overdamped Langevin simulations at 
$T_m=305K$ and $f_m=6pN$. 
The unfolding/refolding time is observed to be between $1$ to $4ms$ (Fig.\ref{barrier_and_time}-B). 
From the free energy profile (Fig.\ref{barrier_and_time}-A) we find $\Delta F^{\ddagger}/T\sim 3$, 
so that $\tau_0=0.05$ to $0.2ms$. 
Consequently, $\tau_h$ at $T=254K$ and $f=12pN$ is 
estimated to be $1$ to $4 s$, which is \emph{three orders of magnitude greater than at the higher $T_m$ and lower $f_m$}.\\

{\bf Thermal refolding and unfolding:}
To induce thermal refolding we performed a temperature quench starting from a
thermally equilibrated ensemble at $T=510K$
to $T=290K<T_m$.
The approach to the folded RNA hairpin is monitored using the time dependence of $Q$, $\chi$, and $n_{bond}$. 
A molecule is in the native state if $Q>0.97$ and $n_{bond}=9.0$.
To confirm that the conformations with these values of $Q$ and $n_{bond}$ are in the NBA we performed steepest descent simulations from states with $Q>0.97$. Most of these conformations 
reach the native state with $\chi=0.00$.   

To calculate the folding time we performed temperature quench simulations for 100 different initially 
denatured conformations to obtain the distribution of the first passage time, 
i.e., the first time molecule $i$ reaches the NBA.
The initial population of unfolded molecules decays exponentially with the folding time $\tau^T_{F}\approx 12.4\mu s$.
Nearly, 90\% of the initially denatured molecules form folded structures in an 
``all-or-none'' manner in which hairpin formation is initiated near the loop region with zipping of 
stabilizing contents progressing towards the end until the 5' and 3' contacts are 
established. 
In rare instance, the 5' and 3' ends meet first and zipping proceeds from the ends to the 
loop region (10\%).
Because of high entropy costs this process occurs is less probable.

For comparison with mechanical unfolding we also performed simulations to monitor thermal unfolding. 
Equilibrated conformations at $T=100K$ are heated to $T=346K$ to initiate
unfolding. 
Unlike in the thermal refolding, in which hairpin is formed by a zipping process, there is no 
characteristic disruption pathway. 
All of the nine bonds fluctuate independently until denaturation occurs. 
Thus, thermal unfolding is stochastic.
Details of thermal unfolding and refolding will be published elsewhere. \\

{\bf Unfolding dynamics at constant force:}
To probe the structural transitions in the hairpin we performed steered Langevin dynamics simulations at constant force at 
$T=254K$. From the phase diagram the equilibrium unfolding force at this temperature 
is $12pN$ (Fig.\ref{phase}-A). 
To monitor the complete unfolding of P5GA, in the time course of the simulations, 
we applied $f=42pN$ to one end of the hairpin with the other end fixed. 
In contrast to thermal unfolding (or refolding) the initially closed hairpin unzips from 
the end to the loop region. 
The unzipping dynamics, monitored by the time dependence of $R$, shows \emph{quantized staircase-like jumps} 
with great variations in step length,
that depends on the initial conditions. 
The lifetimes associated with the ``intermediates'' vary greatly (Fig.\ref{ext_refold}-A).
The large dispersion reflects the heterogeneity of mechanical unfolding pathways.
Approach to the stretched state that occurs in a stepwise ``quantized'' manner \cite{KlimovPNAS99}, which was first
shown in lattice models of proteins, has recently been experimentally observed in the unzipping dynamics of 
DNA under constant force \cite{PrentissPNAS03}. 
The presence of initial condition-dependent unfolding suggests that even in the small P5GA hairpin 
several distinct ``metastable intermediates'' are explored upon stretching.\\

{\bf Refolding under force quench:}
To monitor the dynamics of approach to the NBA we initiated 
refolding from extended conformations with $R=13.5nm$, prepared by 
stretching at $T=290K$ and $f=90pN$. 
Subsequently, we set $f=0$ and the approach to the native state was monitored. 
From the distribution of first passage times 
the refolding kinetics follows exponential kinetics with the mean folding time of 
\emph{about $191\mu s$ compared to $12.4\mu s$ in the temperature quench}.
It is remarkable that, even though the final conditions ($T=290K$ and $f=0$) are the same as in thermal refolding, 
the time scale for hairpin formation $\tau_F^f\approx 15\tau_F^T$!.

The large difference in $\tau_F^T$ and $\tau_F^f$ arises because the molecules under the distinct initial conditions
navigate entirely different regions of the energy landscape. 
The distribution of $R$ in the thermally denatured conformations is 
$P(R)_{thermal}\propto e^{-\beta V_{tot}(R)/k_BT_H}$ ($T_H$ is the initial temperature), 
while in the ensemble of the stretched conformation $P(R)_{stretch}\propto\delta(R-R_{ext})e^{-\beta (V_{tot}(R)-\vec{f}\cdot\vec{R})/k_BT}$. 
The stretched conformations ($R_{ext}=13.5nm$) 
do not overlap with the  the accessible regions of the canonical ensemble of thermally denatured conformations (data not shown). 
As a consequence the regions of the free energy landscape from which folding commences in force jump folding are vastly different 
from those corresponding to the initial population of thermally equilibrated ensemble.
\\

{\bf Force quench refolding occurs in multiple stages:}
The pathways explored by the hairpins en route to the NBA are heterogeneous (Fig.\ref{ext_refold}-B). 
Different molecules reach the hairpin conformation by vastly different routes.
Nevertheless, the time dependence of $R$ shows that the approach to the native conformation occurs in stages (Fig.\ref{ext_refold}-B).
Upon release of force there is a rapid initial decrease in $R$ that results in the collapse of the hairpin. 
Surprisingly, this process takes on an average several $\mu s$, which is much larger than expectations based on 
theories of collapse kinetics of polymer coils \cite{ThirumJPI,PitardEL98}. 
In the second stage, the hairpin fluctuates in relatively compact state with $R$ 
in the broad range (25-75)\AA\ for prolonged time periods. 
On this greatly varying time scales, which varies considerably depending on the molecules, conformational search occurs among compact structures. 
The final stage is characterized by a further decrease in $R$ that takes the molecules to the NBA. 
The last stage is most cooperative and sudden whereas the first two stages appear to much more continuous (Fig.\ref{ext_refold}-B). 
Interestingly, similar relaxation patterns characterized by heterogeneous pathways and continuous collapse in the early stages has been 
observed in force quench refolding of ubiquitin \cite{FernandezSCI04}. 
The multistage approach to the native stage is 
reminiscent of the Camacho-Thirumalai proposal for protein refolding \cite{CamachoPNAS93}. 
\\

\section{Conclusion}
Use of constant force to unfold or initiate refolding (by force quench) provides glimpses of regions of the energy landscape 
of biomolecules that cannot be probed by conventional methods. 
In the mechanical unfolding experiments the molecules go from an initial low entropy state (folded) to another low entropy 
state (stretched). 
This is different from conventional experiments in which unfolding results in a transition from a 
low entropy state to a high entropy state (unfolded). 
This difference results in vastly different mechanisms and time scales of folding and unfolding. 
Using novel coarse-grained models of RNA we have highlighted some of the major differences by considering temperature and 
force effects on unfolding RNA hairpins. 

Our studies have lead to the following predictions, all of which are amenable to experimental tests:
(1) The hairpin undergoes a first order transition from the folded to unfolded states at a critical value of $f$. 
The transition becomes strongly first order at low temperatures and high forces. 
Force unfolding, at a fixed $f$, is more cooperative than unfolding with $T$ fixed and $f$ being varied. 
(2) Unfolding of RNA occurs in steps with long pauses in a number of discrete intermediates that have a large dispersion in $R$ values. 
(3) There are great variations in the hopping times between the NBA and the UBA along the locus of points in the 
($f,T$) plane separating NBA and UBA. 
At low $T_m$ and high $f_m$ the hopping times are orders of magnitude greater than at $T\approx T_m$ and low $f_m$. 
(4) Remarkably, refolding times by force quench are much greater than folding initiated by temperature quench. 
The approach to the native state from stretched conformations occurs in several stages. 
The earliest events involve continuous changes in the progress variable that monitors folding rather being an ``all-or-none'' process. 

\section{Acknowledgements} This work was supported in part by a grant from the National Science Foundation through grant number 
NSF CHE02-09340.
\newpage

\newpage
\section*{\bf Figure Caption}

\begin{figure}[ht]
\includegraphics[width=6.00in]{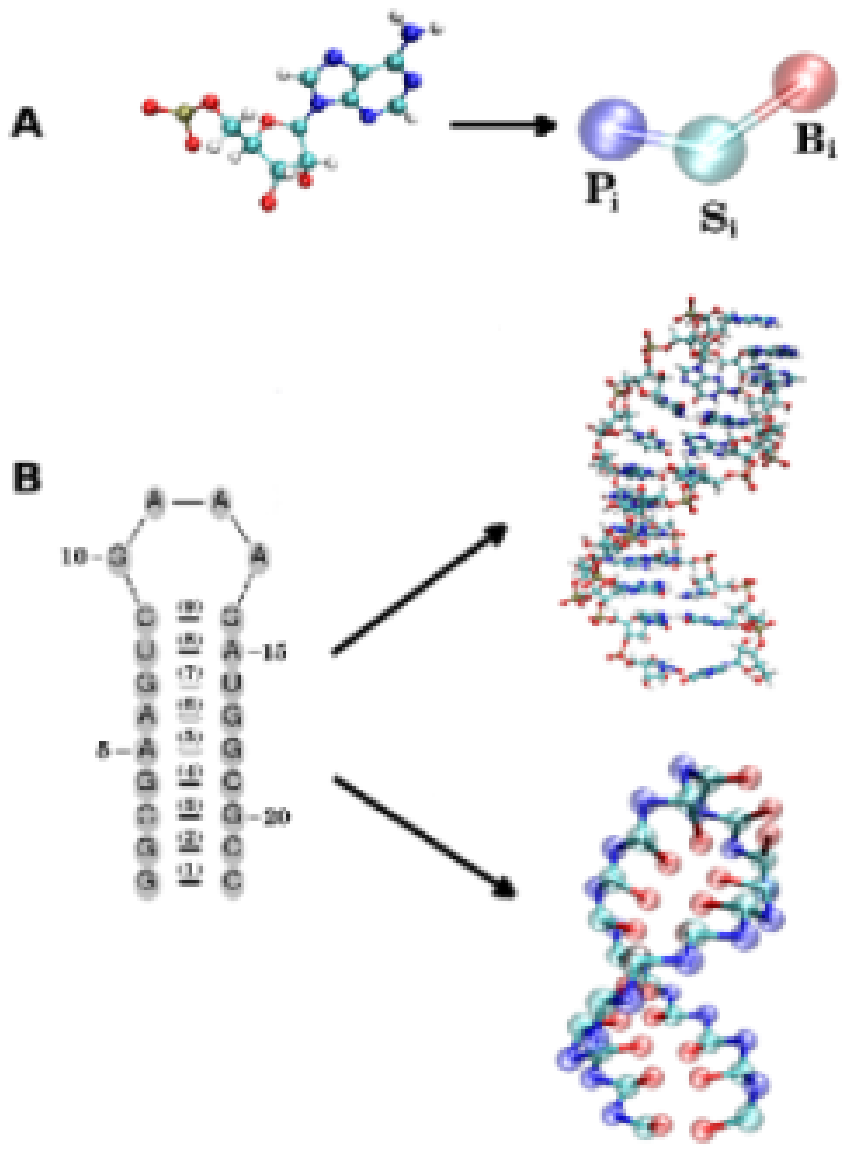}
\caption{\label{coarse_nucl}}
\end{figure}

{{\bf Figure} \ref{coarse_nucl} :}
{\bf A.} Coarse-grained representation of a nucleotide using three sites, namely, phosphate (P), sugar (S) and base (B) are given. 
{\bf B.} The secondary structure of the 22-nt P5GA hairpin in which the bonds formed 
between base pairs are labeled from 1 to 9. 
The PDB structure \cite{TinocoJMB2000} and the corresponding structure using the coarse grained model are shown on the right. 

\begin{figure}[ht]
\includegraphics[width=4.50in]{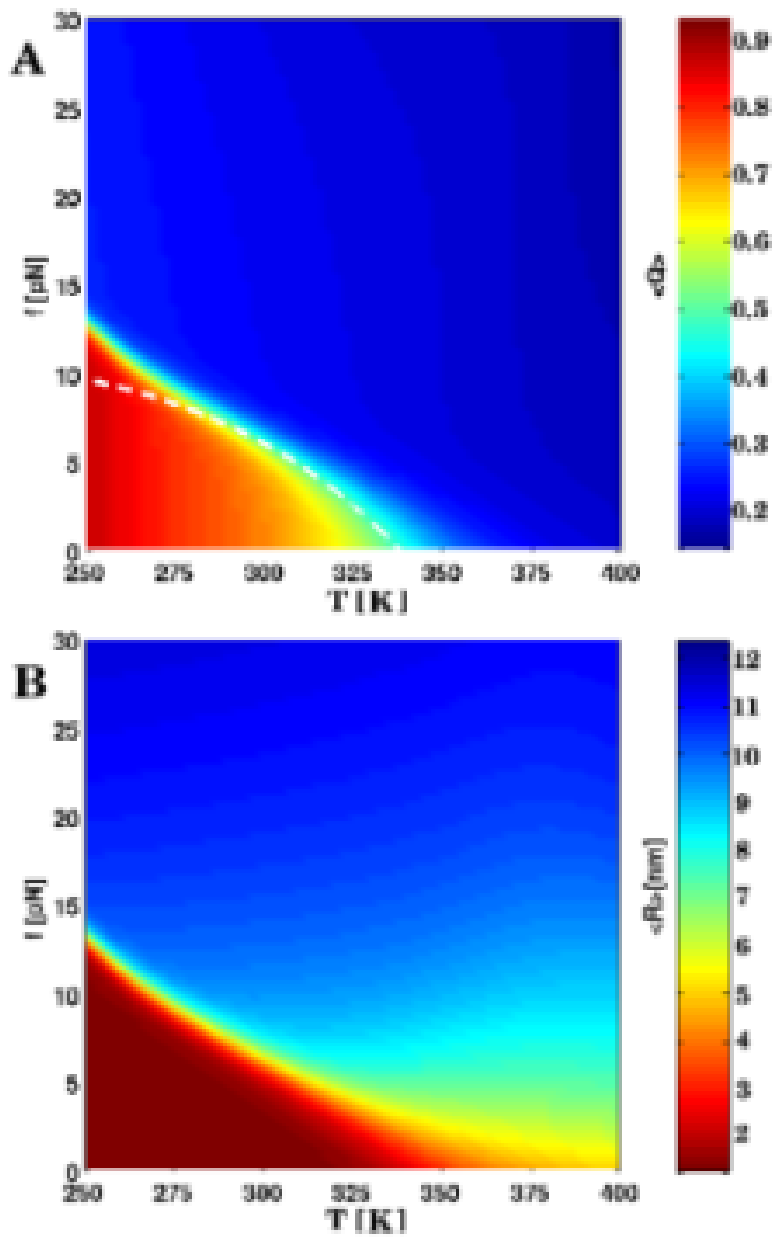}
\caption{\label{phase}}
\end{figure}

{{\bf Figure} \ref{phase} :} 
Phase diagram for the P5GA hairpin. {\bf A.} This panel shows the diagram of states obtained using the 
fraction of native contacts as the order parameter. 
The values of the thermal average of the fraction of native contacts, $\langle Q\rangle$, are color coded as indicated on the scale shown on the right. 
The dashed line is a fit using Eq.(\ref{eqn:boundaryfit}) to the locus of points in the ($f,T$) plane 
that separates the folded hairpin from the unfolded states. 
{\bf B.} Plot of the phase diagram in the ($f,T$) plane using the mean end-to-end distance $\langle R\rangle$ 
as the order parameter. 
Although the diagram of states is qualitatively similar as in {\bf A} there are quantitative differences in estimates of 
$T_m$ at $f=0$. However, estimates of threshold force values at $T<T_m$ are similar in both {\bf A} and {\bf B}.

\begin{turnpage}
\begin{figure}[ht]
\includegraphics[width=8.50in]{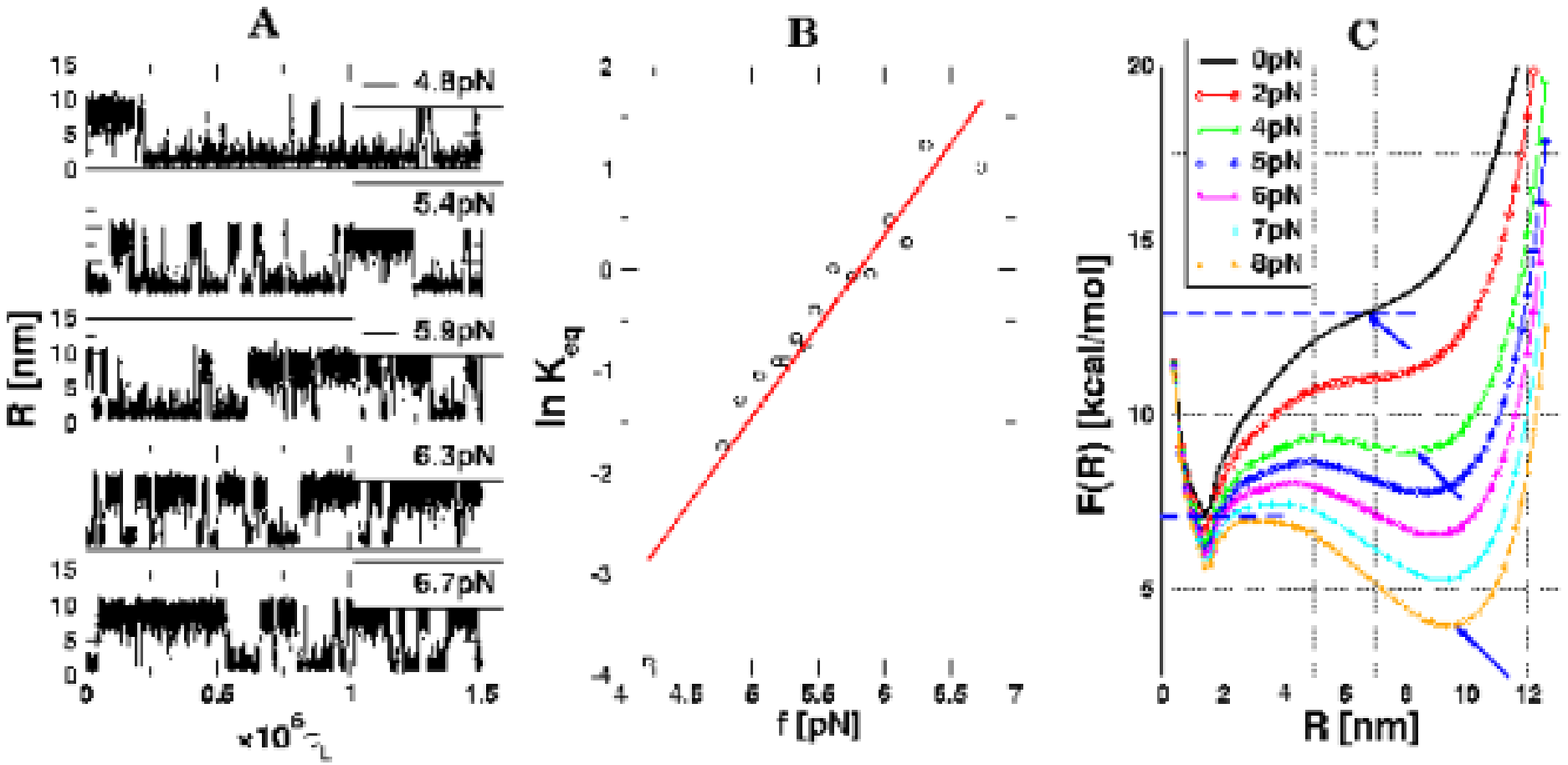}
\caption{\label{compKeq}}
\end{figure}
\end{turnpage}

{{\bf Figure} \ref{compKeq} :}
{\bf A.}  Time traces of $R$ at various values of constant force at $T=305K$.
At $f=4.8pN<f_m\approx 6pN$ $\langle R\rangle$ fluctuates around at low values which shows that 
the NBA is preferentially populated (first panel).  
As $f\sim f_m$ (third panel) the hairpin hops between the folded state (low $R$ value) and unfolded states ($R\approx 10nm$).
The transitions occur over a short time interval. 
These time traces are similar to that seen in Fig.2-C of \cite{Bustamante2}. 
{\bf B.} Logarithm of the equilibrium constant $K_{eq}$ (computed using the time traces in {\bf A}) 
as a function of $f$. 
The red line is a fit with $\log{K_{eq}}=10.4+1.79 f$. 
{\bf C.} Equilibrium free energy profiles $F(R)$ as a function of $R$ at $T=305K$.
The colors represent different $f$ values that are displayed in the inset. The arrows give the location of the unfolded basin of 
attraction.

\begin{figure}[ht]
\includegraphics[width=6.00in]{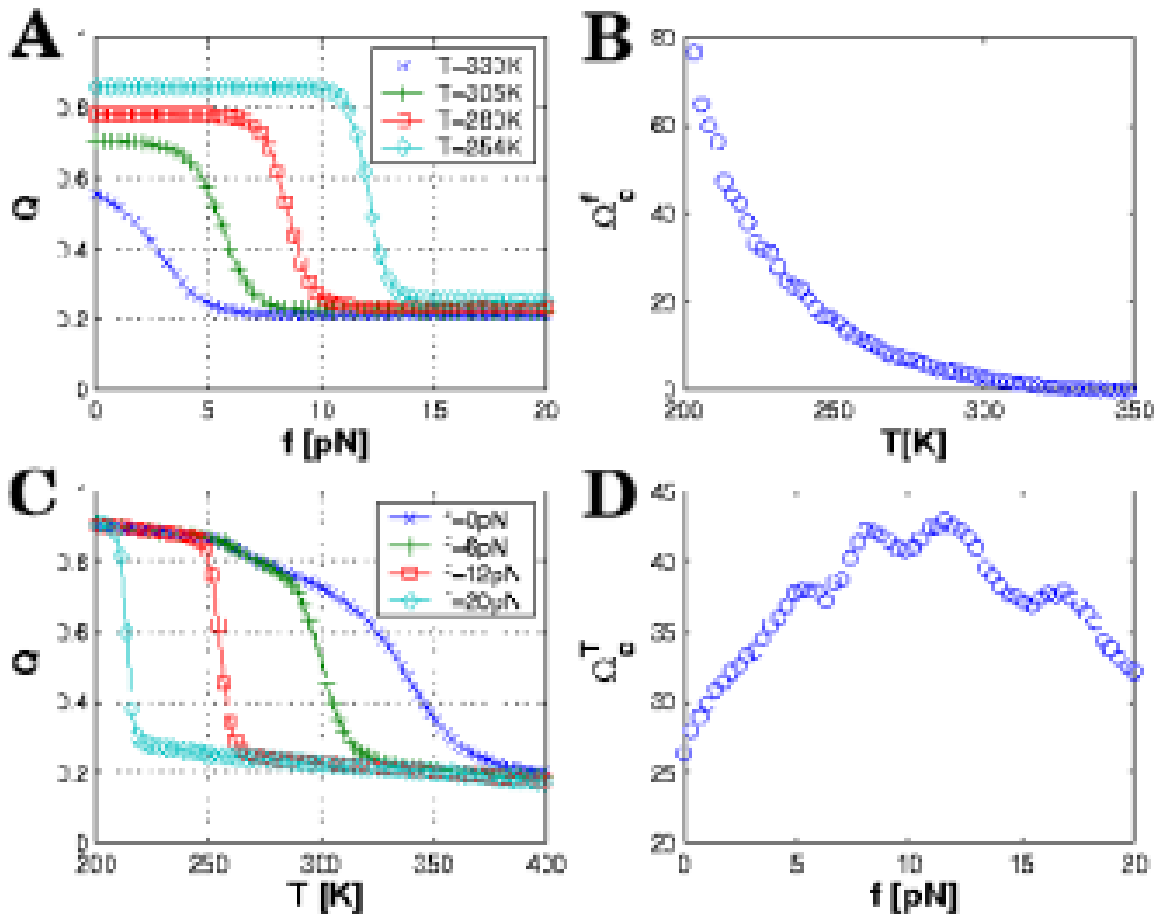}
\caption{\label{coop}}
\end{figure}

{{\bf Figure} \ref{coop} :}
{\bf A.} Dependence of $\langle Q(T,f)\rangle$ as a function of $f$ at various temperatures.
{\bf B.} Values of $(\Omega_c^f)_T$ as a function of temperature.
{\bf C.} Variation of $\langle Q(T,f)\rangle$ as a function of $T$ at various values of $f$. 
{\bf D.} Dimensionless cooperativity measure $(\Omega_c^T)_f$ for $0\leq f\leq20$.

\begin{figure}[ht]
\includegraphics[width=6.00in]{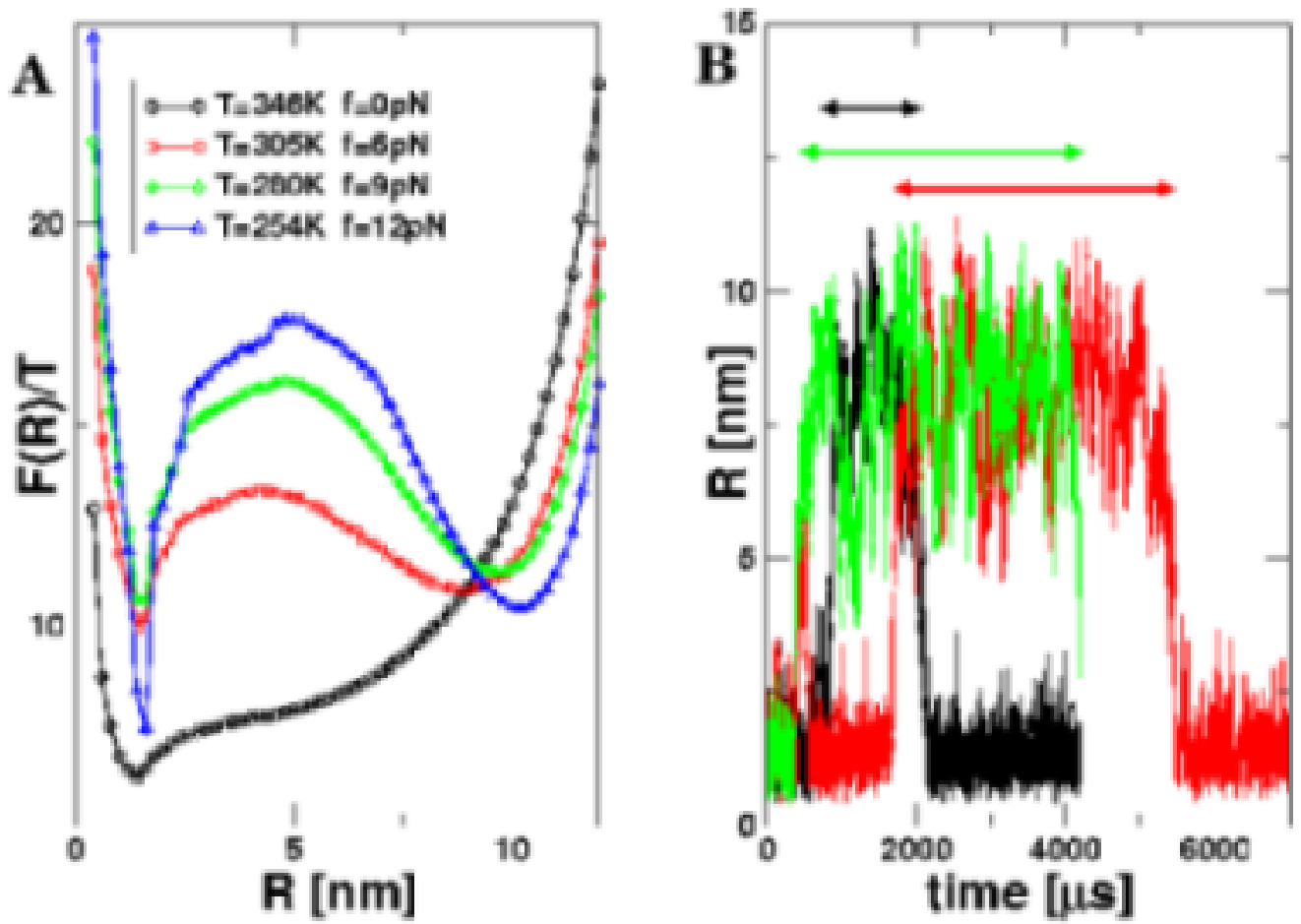}
\caption{\label{barrier_and_time}}
\end{figure}

{{\bf Figure} \ref{barrier_and_time} :}
{\bf A.} Free energy profiles $F(R)$ along the phase boundary ($T_m$,$f_m$) (see Fig.\ref{phase}).
The barrier separating NBA and UBA increases at low $T_m$ and high $f_m$ values. 
{\bf B.} Time traces of $R$ obtained using Brownian dynamics simulations. 
The values of $T$ and $f$ are $305K$ and $6pN$, respectively.  
The arrows (black, red, green) indicate the residence times in the NBA for three trajectories.   

\begin{figure}[ht]
\includegraphics[width=5.00in]{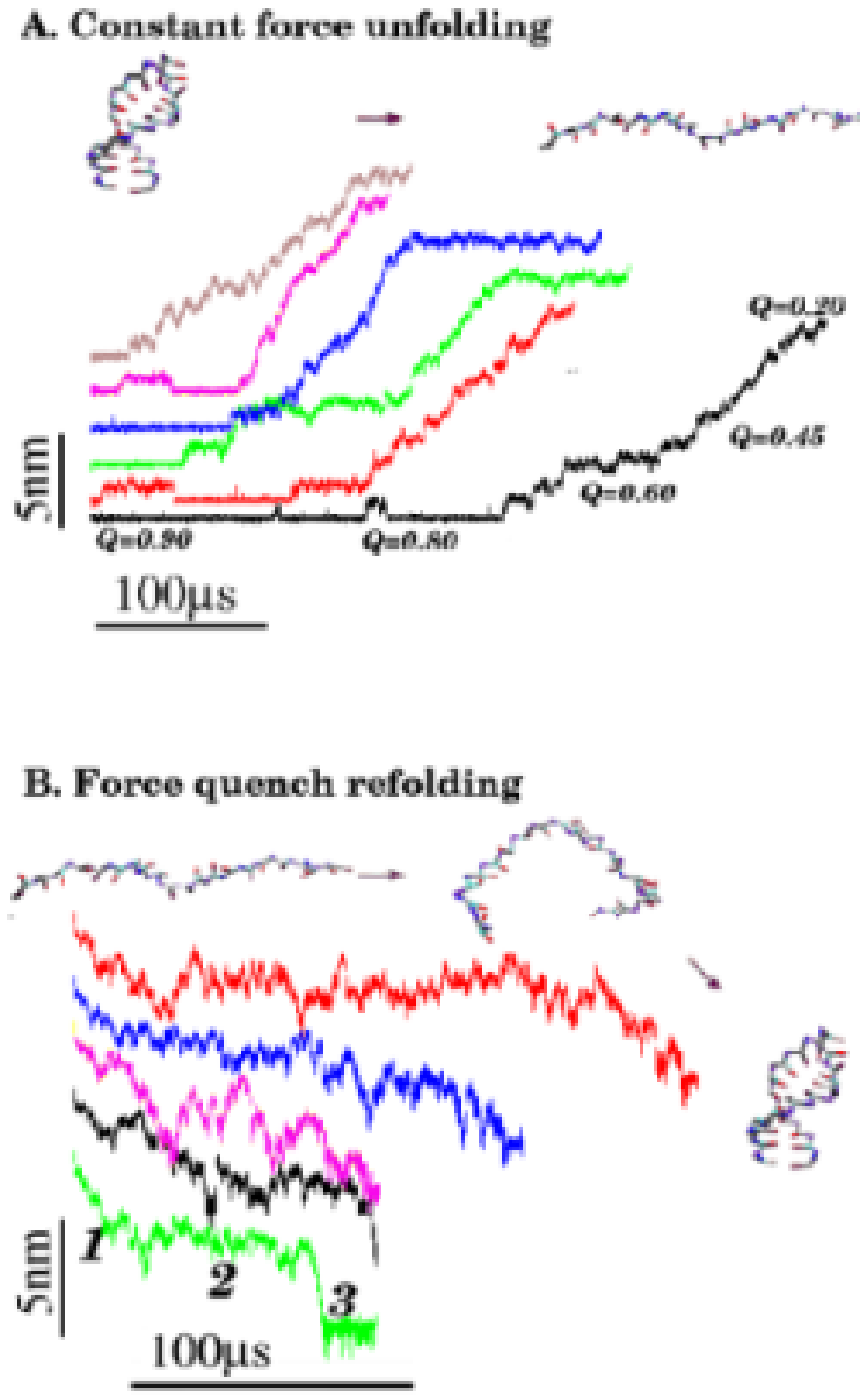}
\caption{\label{ext_refold}}
\end{figure}

{{\bf figure} \ref{ext_refold} :}
{\bf A.} Time traces of unfolding of P5GA at a constant force f=42pN at T=254K monitored by the increase in $R$. The values of $Q$ at different unfolding stages are given for the trjectory in black. 
{\bf B.} Refolding is initiated by a force quench from the initial value f=90pN to f=0. 
The five time traces show great variations in the relaxation to the hairpin conformation. 
However, in all trajectories $R$ decreases in at least three stages that are explicitly labeled for the trajectory in green. 
The trajectories in {\bf A} and {\bf B} are offset for charity.

\newpage
\newpage
\newpage
\newpage
\newpage
\newpage
\end{document}